# Huygens-Fresnel Picture for Electron-Molecule Elastic Scattering

A S Baltenkov[1] and A Z Msezane[2]

[1] Arifov Institute of Ion-Plasma and Laser Technologies,
100125, Tashkent, Uzbekistan
[2] Center for Theoretical Studies of Physical Systems,
Clark Atlanta University, Atlanta, Georgia 30314, USA

**Abstract**
The elastic scattering cross sections for a slow electron by $C_2$ and $H_2$ molecules have been calculated within the framework of the non-overlapping atomic potential model. For the amplitudes of the multiple electron scattering by a target the wave function of the molecular continuum is represented as a combination of a plane wave and two spherical waves generated by the centers of atomic spheres. This wave function obeys the Huygens-Fresnel principle according to which the electron wave scattering by a system of two centers is accompanied by generation of two spherical waves; their interaction creates a diffraction pattern far from the target. Each of the Huygens waves, in turn, is a superposition of the partial spherical waves with different orbital angular momenta $l$ and their projections $m$. The amplitudes of these partial waves are defined by the corresponding phases of electron elastic scattering by an isolated atomic potential. In numerical calculations the $s$- and $p$-phase shifts are taken into account. So the number of interfering electron waves is equal to eight: two of which are the $s$-type waves and the remaining six waves are of the $p$-type with different $m$ values. The calculation of the scattering amplitudes in closed form (rather than in the form of S-matrix expansion) is reduced to solving a system of eight inhomogeneous algebraic equations. The differential and total cross sections of electron scattering by fixed-in-space molecules and randomly oriented ones have been calculated as well. We conclude by discussing the special features of the S-matrix method for the case of arbitrary non-spherical potentials.

## 1. Introduction

The multiple scattering (MS) methodology is one of the most popular theoretical construction for calculating molecular continuum wave functions. The general ideas of this method were developed in paper [1] where "the multiple scattering technique for treating nonseparable eigenvalue problems with electron-scattering theory to construct continuum wave functions" was combined. The MS methodology [1] and its subsequent modifications [2-9] are nowadays widely used to calculate the cross sections for electron elastic scattering by molecules and molecular photoionization. Originally the MS method was used in molecular physics to calculate bound state eigenvalues [10]. For the calculation of the bound state wave functions their normalization is evident. The situation with the continuum wave functions is quite different. A choice of their normalization, i.e. the asymptotic behavior of the wave functions, requires careful analysis. This is of great importance for the accuracy of any method of molecular continuum calculations, particularly when one deals with differential cross sections of fixed-in-space targets, because these cross sections are extremely sensitive to the asymptotic behavior of the wave function.

Within the framework of the approach [1] the potential of the multi-atomic system is represented as a cluster of non-overlapping spherical potentials centered on the atomic sites. In the space between the atomic spheres the potential is assumed to be a constant. This so-called muffin-tin potential (MTP) is the theoretical construction used in solid-state physics where the MTP covers all space. Therefore, there is no question about the potential behavior beyond the microscopic body [11]. However, for the molecular case the situation is quite different. MTP here is created by a finite number of atomic spheres; consequently it is impossible to neglect the existence of the molecular boundary. The adaptation of MTP for this case consists in introducing a *molecular sphere* [1, 2] that surrounds all atoms of the molecule. One of the general ideas of the methods [1-9] is based on an assumption that the radial parts of the electron wave functions outside the molecular sphere with radius $R_m$ can be represented as a linear combination of the regular and irregular solutions of the Schrödinger equation with the proviso that beyond the molecular sphere the potential "is taken to be spherical … about the molecular center" [1]. The molecular phases of scattering are defined by the matching conditions of the continuum wave function on the surfaces of the atomic and molecular spheres. In this scattering picture (see Fig. 1*c* in paper [12]) the wave function beyond the molecular sphere is a sum of a plane wave plus a single spherical wave (SSW) emitted by the molecular center. Hence, in the method [1] the solution of the problem of electron scattering by a non-spherical potential is reduced to the usual S-method of the partial waves for a spherical target.

The general problem of multiple wave scattering by a system of scatterers was investigated long before appearance of the above mentioned papers. The review of these studies is given in Refs. [13-19]. A classical physical picture of wave scattering is based on the Huygens-Fresnel principle, according to which the initial wave interacting with each target center becomes a source of the secondary spherical scattered waves. Therefore, beyond the target there is a system of spherical waves diverging from each of the centers (see Fig. 1*a* in paper [12]), rather than a single spherical wave as proposed in the method [1]. It is known that the interference of the spherical waves emitted by the spatially separated sources creates a diffraction pattern whose properties depend periodically on the ratio of the inter-nuclear distance to the electron wavelength. If we assume, as is done in [1], that beyond the system of the scattering centers there is a single spherical wave, then the phenomenon of electron diffraction by molecules as the interference of a few spherical waves becomes impossible. Departure from the Huygens-Fresnel picture of scattering is equivalent, *per se*, to replacement of diffraction by a system of non-overlapping potentials with electron wave diffraction by a single molecular sphere. Therefore it is difficult to accept that the SSW assumption in [1] can be the basis for the correct description of molecular continuum wave functions.

According to paper [20], in the problem of electron scattering by a cluster of non-overlapping atomic potentials, the wave function corresponding to the Huygens-Fresnel picture of scattering can be represented as a plane wave and a linear combination of the Green functions and the derivatives of these functions. Beyond the atomic spheres the so-written continuum wave function is the accurate solution of the wave equation. The boundary conditions imposed on this function at the centers of the atomic potentials, result in a system of inhomogeneous algebraic equations. Their solution defines the coefficients of the linear combination and the amplitude of electron scattering by a target. Thus, in [20] (see also review [21] and references therein) it was shown that if one knows the elastic scattering phases for each of the spherical potentials forming the target and the target geometry, then the amplitude of elastic scattering can be written in closed form rather than in the form of partial-wave expansion [1].

Note that in nuclear physics a similar method was used by Brueckner [18] for consideration of multiple scattering by two center targets where the wave function of meson scattering by a deuteron was constructed as a plane wave and combination of the Green functions (*s*-spherical waves) and its gradient (*p*-spherical waves). There it was shown that the consideration of the spherical *p*-type wave leads to significant corrections in the scattering cross section as compared to the result of *s*-scattering calculation. In paper [20] a systematic method of building the continuum wave function with *s*, *p*, *d… etc* orbital angular momentum of spherical waves was described.

In the present paper we demonstrate a general methodology based on the equations of paper [20] for the calculation of cross sections for elastic scattering of slow electrons by molecular $C_2$ and $H_2$ within the framework of the model of two non-overlapping atomic spheres. In Sections 2-4 the general formulas for the electron elastic scattering amplitudes are derived. They are used in Section 5 for the numerical calculations of the differential and total cross sections: for the molecules fixed-in-space and randomly oriented molecules in space. In Section 6 the S-matrix method for non-spherical targets is briefly discussed. Section 7 presents the Conclusions. The Appendix A contains some intricate computations. Atomic units are used throughout the paper.

## 2. General formulas for elastic electron scattering by two non-overlapping atomic potentials

Let us consider the elastic scattering of a slow electron by two identical non-overlapping spherically symmetric atomic potentials with the centers at points $\mathbf{r} = \pm \mathbf{R}/2$. We represent the phases of electron elastic scattering by $\delta_l(k)$ off an isolated atomic potential (**k** is the electron wave vector). Following the ideas developed in [20], we represent the wave function of electron scattering by a target beyond the range of atomic potentials *d* in the form of a sum of a plane wave and two sets of *s*-, *p*-, *d*-... spherical *lm*-waves with the centers at points $\mathbf{r} = \pm \mathbf{R}/2$ (see Eq. 12 in [20])

$$\psi_{\mathbf{k}}^{+}(\mathbf{r}) = e^{i\mathbf{k}\cdot\mathbf{r}} + \sum_{l,m} D_{lm}^{+} P_{kl}(\rho_{+}) Y_{lm}(\boldsymbol{\rho}_{+}) + \sum_{l,m} D_{lm}^{-} P_{kl}(\rho_{-}) Y_{lm}(\boldsymbol{\rho}_{-}). \qquad (1)$$

Here the vectors $\boldsymbol{\rho}_{\pm} = \mathbf{r} \pm \mathbf{R}/2$, the spherical functions $Y_{lm}(\boldsymbol{\rho}) \equiv Y_{lm}(\vartheta, \varphi)$ where $\vartheta$ and $\varphi$ are the spherical angles of the vector $\boldsymbol{\rho}$. The radial parts of the spherical waves in Eq. (1) are defined in the usual way [22]

$$P_{kl}(r) = (-1)^l \left(\frac{r}{k}\right)^l \left(\frac{1}{r}\frac{d}{dr}\right)^l \frac{e^{ikr}}{r} \qquad (2)$$

As shown in [20], the spherical $lm$-waves in Eq. (1) can be obtained through the operation of the differential operators $\hat{B}_{lm}^{\pm}$ on the free-particle Green function. The explicit form of these operators is given in the Appendix of [20] and in Appendix A of this paper. With their help the wave function, Eq. (1), can be rewritten as

$$\psi_{\mathbf{k}}^{+}(\mathbf{r}) = e^{i\mathbf{k}\cdot\mathbf{r}} + \sum_{l,m} D_{lm}^{+} \hat{B}_{lm}^{+} G_k(\mathbf{r}, -\mathbf{R}/2) + \sum_{l,m} D_{lm}^{-} \hat{B}_{lm}^{-} G_k(\mathbf{r}, +\mathbf{R}/2). \qquad (3)$$

The coefficients $D_{lm}^{\pm}$ in Eqs. (1) and (3) defining the amplitudes of spherical waves emitted by the centers are calculated from the boundary conditions imposed on the function (1) at the centers of atomic spheres. The number of the spherical $lm$-waves in this wave function is formally equal to infinity. However, the partial wave expansion for slow electron converges rapidly so that the summation usually takes into account only the first few spherical waves with orbital angular momentum $0 \leq l \leq l_{max}$. Each value of the orbital angular momentum $l$ corresponds to $2l+1$ spherical waves, each with different magnetic quantum number $m$. Therefore the number of $lm$-waves emitted from the first center is equal to $(l_{max}+1)^2$. The same number of the waves is emitted by the second center. Hence, the number of coefficients $D_{lm}^{\pm}$ defining the amplitudes of the spherical waves in the wave function, Eq. (3) is equal to $2(l_{max}+1)^2$. A system of the equations for these coefficients has the following form (see Eqs. (24) and (25) in [20])

$$D_{lm}^{+} - 2ikf_l \sum_{\lambda=0}^{l_{max}} \sum_{\mu=-\lambda}^{\lambda} D_{\lambda\mu}^{-} \hat{B}_{\lambda\mu}^{-}[h_l(kR) Y_{lm}^{*}(+\mathbf{R})] = S_{lm}^{+}, \qquad (4)$$

$$D_{lm}^{-} - 2ikf_l \sum_{\lambda=0}^{l_{max}} \sum_{\mu=-\lambda}^{\lambda} D_{\lambda\mu}^{+} \hat{B}_{\lambda\mu}^{+}[h_l(kR) Y_{lm}^{*}(-\mathbf{R})] = S_{lm}^{-}. \qquad (5)$$

where $f_l = 1/[k(\cot\delta_l - i)]$ is the partial wave amplitude of electron scattering by the single atomic potential; $h_l(x) = j_l(x) + in_l(x)$ is the spherical Bessel function of the third kind [23, 24]. The right side terms $S_{lm}^{+}$ and $S_{lm}^{-}$ are described by the following formulas

$$S_{lm}^{+} = 4\pi i^l f_l e^{-i\mathbf{k}\cdot\frac{\mathbf{R}}{2}} Y_{lm}^{*}(\mathbf{k}); \qquad S_{lm}^{-} = 4\pi i^l f_l e^{+i\mathbf{k}\cdot\frac{\mathbf{R}}{2}} Y_{lm}^{*}(\mathbf{k}). \qquad (6)$$

Note that when the $l$-th phase shift generated by the atomic potentials goes to zero, $\delta_l(k) = 0$, the coefficients at the correspondent spherical waves $D_{lm}^{\pm}$ in the function Eq. (3) according to Eqs. (4) and (5) become equal to zero, as it must be to make physical sense.

The $r \to \infty$ asymptotic behavior of the wave function Eq. (3) defines the elastic electron scattering amplitude by the target (see Eq. (29) in [20])

$$F(\mathbf{k},\mathbf{k}',\mathbf{R}) = \frac{1}{2\pi}\left[\sum_{l,m} D^+_{lm}\hat{B}^+_{lm}e^{i\mathbf{k}'\cdot\mathbf{R}/2} + \sum_{l,m} D^-_{lm}\hat{B}^-_{lm}e^{-i\mathbf{k}'\cdot\mathbf{R}/2}\right]. \qquad (7)$$

This equation is the desired solution of the problem of finding the scattering amplitude for a two-center target in closed form (rather than in the form of partial-wave expansion) for any orbital angular momentum of the scattered particle from each center.

### 3. The case of *s*-wave electron scattering by atomic potential

If we apply the general formulas to the case of *s*-wave electron scattering by each of the atomic potentials and put $l_{\max}=0$ in Eqs. (4) and (5) then the solutions of these equations are the ratios of $\|2\|$ determinants:

$$D^+_{00} = \sqrt{4\pi}\,\frac{\begin{Vmatrix} d^* & a \\ d & b \end{Vmatrix}}{\begin{Vmatrix} b & a \\ a & b \end{Vmatrix}}\,;\qquad D^-_{00} = \sqrt{4\pi}\,\frac{\begin{Vmatrix} b & d^* \\ a & d \end{Vmatrix}}{\begin{Vmatrix} b & a \\ a & b \end{Vmatrix}}, \qquad (8)$$

where the following designations are used [20]

$$a = \exp(ikR)/R\,;\ d = -\exp(i\mathbf{k}\cdot\mathbf{R}/2)\,;\ b = k(i-\cot\delta_0) = -1/f_0. \qquad (9)$$

Here $\delta_0(k)$ is the *s*-wave phase shift of the electron wave function for scattering by a single atomic potential; $f_0(k)$ is the partial amplitude of elastic scattering by this potential. The amplitude of the slow electron multiple scattering by the target is defined from the Eq. (7) by the following expression [20]

$$F(\mathbf{k},\mathbf{k}',\mathbf{R}) = \frac{2}{a^2-b^2}\{b\cos[(\mathbf{k}-\mathbf{k}')\cdot\frac{\mathbf{R}}{2}] - a\cos[(\mathbf{k}+\mathbf{k}')\cdot\frac{\mathbf{R}}{2}]\}. \qquad (10)$$

### 4. The case of *s*- and *p*- waves electron scattering by atomic potential

If the first two atomic phase shifts $\delta_0(k)$ and $\delta_1(k)$ are taken into consideration ($l_{\max}=1$), then the wave function for electron scattering by the target Eq. (3) is represented as a sum of a plane wave and two *s*- and six *p*-spherical waves with the centers at $\mathbf{r} = \pm\mathbf{R}/2$. Then the number of equations, from Eqs. (4) and (5), defining the unknown coefficients $D^\pm_{lm}$ increases to eight and we have for $D^\pm_{lm}$ the following matrix equation

$$\begin{Vmatrix} 1 & a_{12} & 0 & 0 & 0 & a_{16} & a_{17} & a_{18} \\ a_{21} & 1 & a_{23} & a_{24} & a_{25} & 0 & 0 & 0 \\ 0 & a_{32} & 1 & 0 & 0 & a_{36} & a_{37} & a_{38} \\ 0 & a_{42} & 0 & 1 & 0 & a_{46} & a_{47} & a_{48} \\ 0 & a_{52} & 0 & 0 & 1 & a_{56} & a_{57} & a_{58} \\ a_{61} & 0 & a_{63} & a_{64} & a_{65} & 1 & 0 & 0 \\ a_{71} & 0 & a_{73} & a_{74} & a_{75} & 0 & 1 & 0 \\ a_{81} & 0 & a_{83} & a_{84} & a_{85} & 0 & 0 & 1 \end{Vmatrix} \cdot \begin{Vmatrix} D^+_{00} \\ D^-_{00} \\ D^+_{1-1} \\ D^+_{10} \\ D^+_{11} \\ D^-_{1-1} \\ D^-_{10} \\ D^-_{11} \end{Vmatrix} = \begin{Vmatrix} S^+_{00} \\ S^-_{00} \\ S^+_{1-1} \\ S^+_{10} \\ S^+_{11} \\ S^-_{1-1} \\ S^-_{10} \\ S^-_{11} \end{Vmatrix}. \qquad (11)$$

Here $S^\pm_{lm}$ are the right hand sides of Eqs. (4) and (5). The zero elements of the determinant $\|8\|$ are connected with the absence of scattering processes of particle waves by the center that generates the given wave. Different from zero, the coefficients $a_{ij}$ at unknown $D^{\pm}_{lm}$ in (11) are calculated in the Appendix A of this paper. The explicit form of the equations for the unknown coefficients $D^{\pm}_{lm}$ is as follows

$$D^+_{00} + a_{12}D^-_{00} + a_{16}D^-_{1-1} + a_{17}D^-_{10} + a_{18}D^-_{11} = S^+_{00},$$
$$a_{21}D^+_{00} + D^-_{00} + a_{23}D^+_{1-1} + a_{24}D^+_{10} + a_{25}D^+_{11} = S^-_{00},$$
$$a_{32}D^-_{00} + D^+_{1-1} + a_{36}D^-_{1-1} + a_{37}D^-_{10} + a_{38}D^-_{11} = S^+_{1-1},$$
$$a_{42}D^-_{00} + D^+_{10} + a_{46}D^-_{1-1} + a_{47}D^-_{10} + a_{48}D^-_{11} = S^+_{10},$$
$$a_{52}D^-_{00} + D^+_{11} + a_{56}D^-_{1-1} + a_{57}D^-_{10} + a_{58}D^-_{11} = S^+_{11},$$
$$a_{61}D^+_{00} + a_{63}D^+_{1-1} + a_{64}D^+_{10} + a_{65}D^+_{11} + D^-_{1-1} = S^-_{1-1},$$
$$a_{71}D^+_{00} + a_{73}D^+_{1-1} + a_{74}D^+_{10} + a_{75}D^+_{11} + D^-_{10} = S^-_{10},$$
$$a_{81}D^+_{00} + a_{83}D^+_{1-1} + a_{84}D^+_{10} + a_{85}D^+_{11} + D^-_{11} = S^-_{11}. \qquad (12)$$

We now demonstrate the use of the above formulas in calculating the cross sections for slow electron scattering by the molecules $C_2$ and $H_2$. The numerical solution of equations (12) reduces to the transformation of the appropriate determinants into a triangular form.

## 5. Numerical calculations

### 5.1. Differential cross sections

In further calculations we assume that the inter-atomic distances $R$ in molecules $H_2$ and $C_2$ are as follows: $R_H$=1.401 and $R_C$=2.479 atomic units (au) [25], respectively. The $s$- and $p$-phase shifts of electron elastic scattering by a single H atom were taken from [14]. The triplet and singlet phase shifts [14] are excellently described by the following expressions: $\delta^t_0(k) = \pi - 1.92564k$ and $\delta^s_0(k) = \pi - 5.72682k + 3.62932k^2$. In the case of C atom the Hartree-Fock function $\delta_0(k)$ was calculated with codes [26] and it is described by the following expression $\delta_0(k) \approx 2\pi - 1.912k$. The $p$-phase shifts for small electron energies are connected with the $s$-phase shifts $\delta_0(k)$ by the following relation [27]: $\delta_1(k) = \delta_0(k)(kd)^2/9$, where $d \sim R/2$ is the radius of the atomic sphere. For the $p$-phase to be calculated, in this formula the term with $\pi$ in $\delta_0(k)$ should be omitted and a corresponding sign of the phase should be taken.

In all the calculations of the scattering amplitudes Eqs.(7) and (10) we assume that the wave vector **k** of the incident electron wave coincides with the Z-axis. So, in the spherical functions $Y_{lm}(\mathbf{k}) = Y_{lm}(\vartheta_k, \varphi_k)$ (see the functions $S^{\pm}_{lm}$) the polar and azimuthal angles of the vector **k** are both equal to zero.

Let us first consider the differential cross section when the vector **R** (the axis of molecule $C_2$) with the spherical coordinates ($R$, $\theta_R$, $\varphi_R$) is perpendicular to the vector **k**. The calculated differential cross section for electron elastic scattering by a molecule $C_2$ is given by

$$\frac{d\sigma}{d\Omega_{k'}} = |F(\mathbf{k},\mathbf{k'},\mathbf{R})|^2 \qquad (13)$$

For different electron energies $\varepsilon = k^2/2$ the differential cross sections are presented in Fig. 1. In the calculation of the amplitudes (7) and (10) it was assumed that the electron wave vector after scattering is in the same plane with the vectors **k** and **R** ($\theta_R$=90°, $\varphi_R$=0°). The molecular atoms are depicted schematically in the figures by solid circles (see upper left panel). The length of the vector **Q** is the differential cross section (13) in the given direction. In this set of figures we can study how the angular distribution of the scattered electron is transformed when its momentum $k$ changes within the range 0.4-0.85 au. With increase in electron energy the spectrum of the angular distribution in the considered plane changes from almost a circular form (for $k$=0.4 au) to a two-leafs one (for $k$=0.85 au). If in calculations only the $s$-phase shift $\delta_0(k)$ is taken into account, then only the first two equations in (12) remain and the scattering amplitude is defined by formula (10). Since $\mathbf{k} \perp \mathbf{R}$ the scattering amplitude (10) reduces to the form

$$F(\mathbf{k},\mathbf{k}',\mathbf{R}) = -\frac{2}{a+b}\cos(\mathbf{k}'\cdot\mathbf{R}/2). \qquad (14)$$

In this case, according to (14) the differential cross section (13) is symmetrical relative to forward/back scattering. Therefore, the dashed curves in Fig. 1 are symmetrical relative to the X-axis. However, the inclusion of *s*- and *p*-phase shifts $\delta_0(k)$ and $\delta_1(k)$ in Eq.(7) leads to the violation of this symmetry. The forward cross section (along the direction of the incident wave vector **k**) represented by the solid curves in Fig. 1, is larger than the back-scattering cross section.

Fig. 2 represents the study of the evolution of the scattering spectra when the angle between the vectors **k** and **R** changes; as before, the electron wave vector after scattering **k'** is in the same plane as with these vectors (k=0.8 au). As in Fig. 1, the C atoms of the molecule are depicted schematically by the solid circles. For the angle $\theta_R=90°$ the spectrum on the lower right panel corresponding to $\theta_R=75°$ is reduced to that on the lower left panel in Fig. 1. Since the scalar product of the vectors **k** and **R** in (10) in this case is not equal to zero, the dashed curves (that corresponds to *s*-scattering by the C-atoms only) stop being mirror-symmetrical relative to the X-axis.

The azimuthal-dependence of the spectra is presented in Fig. 3. Here the vectors **k** and **k'** as before are located in the plane of the page (k=0.85 au) but the vector **R** is not in this plane. The polar angle of this vector is constant with $\theta_R=90°$, while the azimuthal angle $\varphi_R$ has the values 30°, 45°, 60° and 90°. As seen from Fig. 3, with the increase in azimuthal angle $\varphi_R$ the spectrum loses its "waist" along the X-axis and in the case $\varphi_R=90°$ (lower right panel) the dashed curve transforms to a circle as follows from Eq.(14), where owing to the orthogonality of the vectors **k'** and **R,** the cosine in Eq.(14) becomes equal to unity.

## 5.2. Total cross sections

The total scattering cross sections are obtained from the amplitudes (7) and (10) with the help of the optical theorem [22]

$$\sigma(\mathbf{k},\mathbf{R}) = \int \frac{d\sigma}{d\Omega_{k'}} d\Omega_{k'} = \frac{4\pi}{k}\operatorname{Im} F(\mathbf{k}=\mathbf{k}',\mathbf{R}). \qquad (15)$$

We introduce the vectors **k** and **R** in the argument of the cross section (15) to underscore that we are dealing with the fixed-in-space molecule. In the case of the amplitude (10) we have the following compact formula for the cross section

$$\sigma_s(\mathbf{k},\mathbf{R}) = \frac{8\pi}{k}\operatorname{Im}\left[\frac{b-a\cos(\mathbf{k}\cdot\mathbf{R})}{a^2-b^2}\right]. \qquad (16)$$

For the case of the amplitude (7) the total cross section is described by the following expression

$$\sigma_{sp}(\mathbf{k},\mathbf{R}) = \frac{2}{k}\operatorname{Im}\sum_{l=0}^{1}\sum_{m=-1}^{1}\left[D_{lm}^{+}\hat{B}_{lm}^{+}e^{i\mathbf{k}\cdot\mathbf{R}/2} + D_{lm}^{-}\hat{B}_{lm}^{-}e^{-i\mathbf{k}\cdot\mathbf{R}/2}\right]. \qquad (17)$$

The total cross section (16) averaged over all the directions of the incident electron momentum **k** relative to the vector **R** has the form:

$$\bar{\sigma}(k) = \frac{1}{4\pi}\int \sigma(k,\mathbf{R})d\Omega_k = \frac{8\pi}{k}\operatorname{Im}\left[\frac{b-aj_0(kR)}{a^2-b^2}\right]. \qquad (18)$$

Here $j_0(x) = \sin x / x$ is the spherical Bessel function [23]. Using the explicit expressions (9) for the functions $a$ and $b$, we obtain the following formula for the averaged cross section

$$\bar{\sigma}(k) = \frac{4\pi}{k^2} \left\{ \left[ 1 + \left( \frac{qR + \cos kR}{kR + \sin kR} \right)^2 \right]^{-1} + \left[ 1 + \left( \frac{qR - \cos kR}{kR - \sin kR} \right)^2 \right]^{-1} \right\}, \qquad (19)$$

where the wave vector $q = -k \cot \delta_0(k)$. The total cross section (19) contains the term $\sin kR / kR$ characteristic of diffraction phenomena. Its appearance in this case is connected with the interference of two *s*-waves in the continuum wave function (1). Fig. 4 is a vivid illustration of the diffraction character of electron wave scattering by $H_2$ and $C_2$ molecules. The two upper sets of curves in this figure correspond to electron *s*-wave scattering by H and C atomic spheres of the corresponding molecules. The first maximum on the $C_2$ curves corresponds to the wave vector $k \approx 0.6$. The maximum of the $H_2$ curves is located at $k \approx 1.2$; this is because the inter-atomic distance in the $C_2$ molecule is almost two-times greater than that in the $H_2$ molecule. That is, the Huygens-Fresnel diffraction by two-atomic molecules is defined by the product $kR$ where $R$ is the inter-atomic distance rather than the product $kR_m$ where $R_m$ is the radius of the molecular sphere as in the methodology [1]. The middle set of curves in Fig. 4 corresponds to the interference of eight spherical waves of *s*- and *p*-types generated by two hydrogen atoms of $H_2$ molecule. Here we also observe the diffraction pattern, but the alternation between maxima and minima in these curves is not so clearly associated with the ratio of electron wavelength to inter-atomic distance in the molecule.

In Fig. 5 the results of our calculations for randomly oriented molecule $H_2$ are given together with some theoretical and experimental data. Dashed red line is our result when only the *s*-phase scattering is taken into account. The solid blue line is the result when the *s+p* phases are taken into account in the calculation. The discrepancy between our results and experiments for very low electron energies might be, at least in part, due to our neglect of multi-electron correlations. They should have the effect of additional attractive contributions to the electron molecule interaction and could be sufficient for the total cross section to decrease at very low electron energies. In general, the calculated curves are lower than experimental ones, which is possibly associated with insufficient taking into consideration *s+p* phase shifts only. It can be expected that the inclusion in calculations the phase shifts with higher orbital moments will result in increasing the integral cross section of elastic scattering.

Table 2. Experimental data for $\bar{\sigma}(k)$ from [35]

| $\varepsilon$, eV | $\bar{\sigma}(k)$, Å$^2$ |
|---|---|
| 2.5 | 17.1 |
| 3.0 | 16.8 |
| 4.0 | 16.2 |
| 6.0 | 15.4 |
| 8.0 | 12.2 |
| 10.0 | 10.1 |
| 15.0 | 7.17 |

**6. Method of partial waves for non-spherical targets**

The general formulas used above are naturally generalized for the case of an arbitrary cluster of atomic spheres. With these formulas it is possible to obtain the scattering amplitude in a closed form rather than in the form of amplitude expansion in spherical partial waves, as assumed in the method [1]. There the solution of the problem of electron scattering by a spherically non-symmetrical potential was reduced without any justification to the usual method of partial waves for a spherical target. Consequently, a natural question arises: Is it possible to adapt the method of partial waves to the case of non-spherical targets while keeping the Huygens-Fresnel picture of the scattering process? Demkov and Rudakov [32] provided a positive answer to this question by showing that the S-matrix method could be applied to non-spherical potentials as well.

We first describe briefly the main ideas of paper [32]. It is known that the wave function for elastic scattering of a particle by a single spherically symmetrical potential is determined by the expression [22]

$$\psi_{\mathbf{k}}^{+}(\mathbf{r}) = 4\pi \sum_{l=0}^{\infty} R_{klm}(r) Y_{lm}^{*}(\mathbf{k}) Y_{lm}(\mathbf{r}),  \quad (20)$$

where the radial part of the wave function has the asymptotic form

$$R_{klm}(r \to \infty) \approx e^{i(\delta_l + \frac{\pi l}{2})} \frac{1}{kr} \sin(kr - \frac{\pi l}{2} + \delta_l). \quad (21)$$

Following the ideas of the method [1] the molecular continuum wave function is represented in this form beyond the molecular sphere (see for example Eq.(3) in [33]).

A molecular potential as a cluster of non-overlapping spherical potentials centered at the atomic sites is a non-spherical potential. However, in the Schrödinger equation with this potential it is impossible to separate the angular variables and represent the wave function at an arbitrary point of space in the form of an expansion in spherical functions (20). Nevertheless, asymptotically at great distances from the molecule the wave function can be written as an expansion in a set of other orthonormal functions $Z_\lambda(\mathbf{k})$:

$$\psi_{\mathbf{k}}^{+}(\mathbf{r} \to \infty) \approx 4\pi \sum_{\lambda} R_{k\lambda}(r) Z_\lambda^{*}(\mathbf{k}) Z_\lambda(\mathbf{r}) \quad (22)$$

with the radial part of the wave function similar to (21)

$$R_{k\lambda}(r \to \infty) \approx e^{i(\eta_\lambda + \frac{\pi}{2}\omega_\lambda)} \frac{1}{kr} \sin(kr - \frac{\pi}{2}\omega_\lambda + \eta_\lambda). \quad (23)$$

Here the index $\lambda$ enumerates the different partial functions similar to the quantum numbers $l$ and $m$ for the central field; $\omega_\lambda$ is the quantum number that is equal to the orbital angular momentum $l$ for the spherical symmetry case; $\eta_\lambda(k)$ are the molecular phases. The explicit form of the functions $Z_\lambda(\mathbf{k})$, naturally, depends on a specific type of the target field, particularly on the number of atoms forming the target and on the relative orientation of the scattering centers in space, *etc*. The functions $Z_\lambda(\mathbf{k})$, like the spherical functions $Y_{lm}(\mathbf{k})$, create an orthonormal system according to:

$$\int Z_\lambda^{*}(\mathbf{k}) Z_\mu(\mathbf{k}) d\Omega_k = \delta_{\lambda\mu}. \quad (24)$$

The scattering amplitude for a non-spherical target, according to [32], is given by the following expression

$$F(\mathbf{k},\mathbf{k'}) = \frac{2\pi}{ik} \sum_{\lambda} (e^{2i\eta_\lambda} - 1) Z_\lambda^{*}(\mathbf{k}) Z_\lambda(\mathbf{k'}). \quad (25)$$

The total elastic scattering cross section, i.e. the cross section integrated over all directions of the momentum of the scattered electron **k'**, is then defined by the formula

$$\sigma(\mathbf{k}) = \frac{(4\pi)^2}{k^2} \sum_{\lambda} |Z_\lambda(\mathbf{k})|^2 \sin^2 \eta_\lambda. \quad (26)$$

Of course, this cross section depends on the orientation of the incident electron momentum **k** with respect to the molecule axes. The cross section averaged over all the directions of the incident electron momentum **k** is connected with the molecular phases $\eta_\lambda(k)$ through the following formula

$$\bar{\sigma}(k) = \frac{4\pi}{k^2} \sum_\lambda \sin^2 \eta_\lambda . \qquad (27)$$

In the case of a spherically symmetrical target the formula (27) exactly coincides with the well-known formula for the total scattering cross section. Indeed, in the case of the central field the index $\lambda$ is replaced by the quantum numbers *l* and *m*. But the phase of scattering by the central field is independent of the magnetic quantum number and therefore for a given value of the orbital angular momentum *l*, it is necessary to sum over all *m*. This results in the factor (2*l*+1) under the summation sign in formula (27). The partial wave (23) and molecular phases $\eta_\lambda(k)$ are classified, according to [32], by their behavior for low-energy electrons, i.e. for $k \to 0$. In this limit the particle wavelength is great as compared with the target size and the function $Z_\lambda(\mathbf{k})$ tends to some spherical function $Y_{lm}(\mathbf{k})$. The corresponding phase is characterized in this limit by the asymptotic behavior: $\eta_\lambda(k) \to k^{2\lambda+1}$.

For the above-considered molecular system that is created by two short-range potentials each of which being a source of the scattered *s*-waves, the molecular phase shifts $\eta_\lambda(k)$ and the functions $Z_\lambda(\mathbf{k})$ can be calculated explicitly. This simplest multicenter system is a good example for illustrating the method of partial waves for non-spherical targets formed by two non-overlapping atomic potentials. The scattering amplitude (10) that exactly corresponds to the Huygens-Fresnel picture is written in the form

$$F(\mathbf{k},\mathbf{k'},\mathbf{R}) = -\frac{2}{a+b}\cos(\mathbf{k}\cdot\mathbf{R}/2)\cos(\mathbf{k'}\cdot\mathbf{R}/2) + \frac{2}{a-b}\sin(\mathbf{k}\cdot\mathbf{R}/2)\sin(\mathbf{k'}\cdot\mathbf{R}/2) . \qquad (28)$$

According to [32], the amplitude (28) should be considered as a sum of two partial amplitudes. The first of them is written as

$$\frac{4\pi}{2ik}(e^{2i\eta_0}-1)Z_0(\mathbf{k})Z_0^*(\mathbf{k'}) = -\frac{2}{a+b}\cos(\mathbf{k}\cdot\mathbf{R}/2)\cos(\mathbf{k'}\cdot\mathbf{R}/2) . \qquad (29)$$

The second is defined by the following expression

$$\frac{4\pi}{2ik}(e^{2i\eta_1}-1)Z_1(\mathbf{k})Z_1^*(\mathbf{k'}) = \frac{2}{a-b}\sin(\mathbf{k}\cdot\mathbf{R}/2)\sin(\mathbf{k'}\cdot\mathbf{R}/2) . \qquad (30)$$

The reasons for assigning the indices to the functions $Z_\lambda(\mathbf{k})$ the values $\lambda = 0,1$ will become understandable as we proceed. After elementary transformations of the formulas (29) and (30), we obtain two molecular phases of scattering (the proper phases in [32])

$$\cot\eta_0 = \frac{\mathrm{Re}[(a+b)^*]}{\mathrm{Im}[(a+b)^*]} = -\frac{qR+\cos kR}{kR+\sin kR}, \quad \cot\eta_1 = \frac{\mathrm{Re}[(a-b)^*]}{\mathrm{Im}[(a-b)^*]} = -\frac{qR-\cos kR}{kR-\sin kR} . \qquad (31)$$

Substituting the phase shifts (31) into formulas (29) and (30), we obtain the functions $Z_\lambda(\mathbf{k})$ in the explicit form defined by the following expressions ( $Z_\lambda(\mathbf{k})$ are the characteristic scattering amplitudes in [32])

$$Z_0(\mathbf{k}) = \frac{\cos(\mathbf{k}\cdot\mathbf{R}/2)}{\sqrt{2\pi S_+}}, \quad Z_1(\mathbf{k}) = \frac{\sin(\mathbf{k}\cdot\mathbf{R}/2)}{\sqrt{2\pi S_-}} . \qquad (32)$$

Here $S_\pm = 1 \pm j_0(kR)$. It is easy to ensure that the functions (32) obey the conditions (24). Evidently, the functions (32) are defined by the geometrical target structure, i.e. by the direction of the molecular axis **R** in the arbitrary coordinate system in which the electron momentum vectors before and after scattering are **k** and **k'**, respectively.

We now study the asymptotical behavior of the wave functions (22) and (23). Toward this end we write the exponent $\exp(i\mathbf{k}\cdot\mathbf{r})$ in the formula (1) as an expansion in the functions $Z_\lambda(\mathbf{k})$:

$$e^{i\mathbf{k}\cdot\mathbf{r}} = \sum_\lambda c_\lambda Z_\lambda(\mathbf{k}) . \tag{33}$$

Multiplying both sides of this equality by $Z_\mu^*(\mathbf{k})$ and integrating over all angles of the vector **k**, we obtain the following expressions for the coefficients of the expansion (33):

$$c_0 = \sqrt{\frac{2\pi}{S_+}}[j_0(k\,|\mathbf{r}+\mathbf{R}/2|) + j_0(k\,|\mathbf{r}-\mathbf{R}/2|)],$$

$$c_1 = -i\sqrt{\frac{2\pi}{S_-}}[j_0(k\,|\mathbf{r}+\mathbf{R}/2|) - j_0(k\,|\mathbf{r}-\mathbf{R}/2|)] . \tag{34}$$

At large distances from the target the expansion coefficients in equations (34) have the form

$$c_0(r\to\infty) \approx 4\pi\frac{\sin kr}{kr}Z_0^*(\mathbf{k}) \quad \text{and} \quad c_1(r\to\infty) \approx -i4\pi\frac{\cos kr}{kr}Z_1^*(\mathbf{k}) . \tag{35}$$

Consider now the asymptotic behavior of the partial wave with the index $\lambda = 0$. Taking into account the formulas (1), (29), (33) and (35), we write the corresponding partial wave from formula (22) in the form

$$4\pi R_{k0}(r)Z_0(\mathbf{k})Z_0^*(\mathbf{k}) = \frac{4\pi}{kr}[\sin kr + \frac{1}{2i}(e^{2i\eta_0}-1)e^{ikr}]Z_0(\mathbf{k})Z_0^*(\mathbf{k}) . \tag{36}$$

From equation (36) we immediately obtain

$$R_{k0}(r\to\infty) = e^{i\eta_0}\frac{1}{kr}\sin(kr+\eta_0) . \tag{37}$$

Following the same operation for the case $\lambda = 1$, we obtain for the second partial wave the following expression

$$R_{k1}(r\to\infty) = e^{i(\eta_1+\frac{\pi}{2})}\frac{1}{kr}\sin(kr-\frac{\pi}{2}+\eta_1) . \tag{38}$$

The molecular phases $\eta_\lambda(k)$ in (37) and (38) can be classified by considering their behavior as $k\to 0$ [32]. In this limit the electron wavelength is much greater than the target size and the picture of scattering should approach the spherical symmetry one. Consider this limit transition in the formulas (31); we obtain: $\eta_0(k\to 0) \sim k$ and $\eta_1(k\to 0) \sim k^3$. Thus, the molecular phases behave similar to the *s*- and *p*- phases in the spherically symmetrical potential, which explains the choice of their indices. The transition to the limit $k\to 0$ in formulas (32) gives instead of the functions $Z_\lambda(\mathbf{k})$ the well-known spherical functions

$$Z_0(\mathbf{k})_{k\to 0} \to \frac{1}{\sqrt{4\pi}} \equiv Y_{00}(\mathbf{k}), \quad Z_1(\mathbf{k})_{k\to 0} \to \sqrt{\frac{3}{4\pi}}\cos\vartheta \equiv Y_{10}(\mathbf{k}) . \tag{39}$$

Here $\vartheta$ is the angle between the vector **k** and axis **R**.

Finally, substituting the molecular phases (31) in the formula (27), we obtain the following cross section for elastic scattering [32, 34]

$$\bar{\sigma}(k) = \frac{4\pi}{k^2}[\sin^2\eta_0 + \sin^2\eta_1] = \frac{4\pi}{k^2}[(1+\cot^2\eta_0)^{-1} + (1+\cot^2\eta_1)^{-1}]. \qquad (40)$$

The same result was obtained in equation (19), with the help of the optical theorem.

Developed in [32], the theory of partial waves for non-spherical targets separates in a natural way the kinematics of the scattering process (relative orientation of the vectors **k**, **k'** and **R** in the characteristic scattering amplitudes $Z_\lambda(\mathbf{k})$) from the phases of the radial parts of the wave functions (22). The molecular phases $\eta_0(k)$ and $\eta_1(k)$ in (31) depend only on the electron kinetic energy. If one assumes according to the ideas of the method [1] that beyond the molecular sphere the continuum wave function has the form (20) then one comes to the conclusion that the scattering phases in the radial parts of the wave function (23) are the functions of the vector **R**. Indeed, the scattering amplitude (10) corresponding to the function (21) has the form of the amplitude of scattering by a spherically symmetrical potential [22]

$$F(\mathbf{k},\mathbf{k'},\mathbf{R}) = \frac{4\pi}{2ik}\sum_{lm}(e^{2i\eta_l}-1)Y_{lm}(\mathbf{k})Y_{lm}^*(\mathbf{k'}). \qquad (41)$$

The function on the left side of Eq.(41) is a function of three vectors **k**, **k'** and **R**. The spherical functions in (41) cannot depend on the **R** vector. Hence, the **R**-dependence is inside the molecular phase shifts $\eta_l$. Evidently, this physically impossible result is a consequence of the SSW-approach and refusal of the Huygens-Fresnel picture for the molecular continuum wave function.

The possibility of calculating the functions $Z_\lambda(\mathbf{k})$ and the phases $\eta_\lambda(k)$ in explicit forms is connected with the representation of the scattering amplitude (10) in closed form for MTP targets. When the scattering amplitude is known the cross sections can be found with the help of the optical theorem, and therefore it is unnecessary to resort to the method of partial waves. However, for arbitrary non-spherical potentials (different from the muffin-tin-potential) the application of the partial wave method [32] makes it possible to separate explicitly the scattering dynamics contained in the molecular phases $\eta_\lambda(k)$ from the kinematics of the process defined by the functions $Z_\lambda(\mathbf{k})$.

## 7. Conclusions

The multiple scattering of slow electron has been evaluated in the case of two identical atoms representing a simple example of 8 spherical waves in elastic scattering. The calculated spectra of the angular distribution of elastically scattered electrons for a fixed-in-space target are significantly transformed when besides the interference of the two *s*-spherical waves the six spherical waves of *p*-type begin to play a role. The mirror-symmetry of the *s*-spectra relative to the X-axis in Fig. 1 disappears when the *p*-spherical waves are taken into account in the scattering process. With the increase in electron energy a "slender waist" appears in the differential cross sections and the spectrum of the angular distribution acquires a two-leaf form. A strong dependence of the spectrum on polar angle appearing between the vectors **k** and **R** (Fig. 2) has been established. The *s*- and *s*+*p*- spectra follow the vector **R**, significantly decreasing in amplitude. Of great interest is the evolution of the scattering spectra for molecular axis rotation relative to the direction of the incident electron wave vector for fixed polar angle and azimuthal dependence (Fig. 3) as well as that of the total cross section of electron scattering of fixed-in-space and randomly oriented molecules $H_2$ and $C_2$ (Figs. 4 and 5).

Summarizing the results obtained with the S-matrix method [32] for MTP-targets (Section 6), we come to the following conclusion. For the two atomic targets the molecular phases of scattering and the functions $Z_\lambda(\mathbf{k})$ can be found explicitly. The form of the functions $Z_\lambda(\mathbf{k})$ is defined by the structure of a target and its orientation in space. The phases of molecular scattering, as in the case of spherically symmetrical potentials, are functions of the electron momentum $k = |\mathbf{k}|$ only. The number of non-zero molecular phases in the case considered is equal to two. This is connected with the fact that each of the two scattering centers is a source of *s*-spherical waves only, which is valid for the case of low-energy electrons. If the scattering by each of these centers would be accompanied by the generation of spherical waves with non-zero orbital angular momentum $l$, then the number of non-zero molecular phases $\eta_\lambda(k)$ would be greater.

**Acknowledgments**

This work was supported by the Uzbek Foundation Award OT-Ф2-46 (ASB) and U.S. DOE, Basic Energy Sciences, Office of Energy Research (AZM).

**References**

1. D. Dill and J. L. Dehmer, J. Chem. Phys. **61**, 692 (1974).
2. D. Dill and J. L. Dehmer, Phys. Rev. Lett. **35**, 213 (1975).
3. J. Siegel, D. Dill and J. L. Dehmer, J. Chem. Phys. **64**, 3204 (1976).
4. J. W. Davenport, Phys. Rev. Lett. **36**, 945 (1976).
5. M. Venuti, M. Stener, P. Decleva, Chem. Phys. **234**, 95 (1998).
6. P. Downie and I. Powis, Phys. Rev. Lett. **82**, 2864 (1999).
7. P. Decleva, G. De Alti, G. Fronzoni and M. Stener, J. Phys. B **32**, 4523 (1999).
8. Y. Nikosaka, J. H. D. Eland, T. M. Watson and I. Powis, J. Chem. Phys. **115**, 4593 (2001).
9. A. V. Golovin and N. A. Cherepkov, J. Phys. B **35**, 3191 (2002).
10. K. H. Johnson, *Advances in Quantum Chemistry*, edited by P. O. Löwdin (Academic, New York, 1973), vol. **7**, 143.
11. T. L. Loucks, *The Augmented Plane Wave Method* (Benjamin, New York, 1967).
12. A. S. Baltenkov, U. Becker, S. T. Manson and A. Z. Msezane, J. Phys. B **45**, 035202 (2012).
13. M. L. Goldberger and K. M. Watson, *Collision Theory* (John Wiley & Sons, Inc., New York-London-Sydney, 1964).
14. N. F. Mott and H. S. W. Massey, *The Theory of Atomic Collisions* (Clarendon Press, Oxford, 1965).
15. L. L. Foldy, Phys. Rev. **67**, 107 (1945).
16. M. Lax, Rev. Mod. Phys. **23**, 287 (1951).
17. H. Ekstein, Phys. Rev. **87**, 31 (1952).
18. K. A. Brueckner, Phys. Rev. **89**, 834 (1953).
19. K. M. Watson, Phys. Rev. **105**, 1388 (1957).
20. A. S. Baltenkov, S. T. Manson and A. Z. Msezane, J. Phys. B **40** 769 (2007).
21. A. K. Kazansky and I. I. Fabrikant, Physics-Uspekhi, **143** 601 (1984), [in Russian].
22. L. D. Landau and E. M. Lifshitz, *Quantum Mechanics, Non-Relativistic Theory* (Pergamon Press, Oxford, 1965).
23. M. Abramowitz and I. A. Stegun, *Handbook of Mathematical Functions* (New York: Dover, 1965).
24. D. A. Varshalovich, A. N. Moskalev, V. K. Khersonskii, *Quantum Theory of Angular Momentum* (World Scientific, Singapore-New Jersey-Hong Kong, 1988).
25. B. M. Smirnov, *Atomic Collisions and Processes in Plasma* [in Russian] (Atomizdat, Moscow, 1968).
26. M. Ya. Amusia and L.V. Chernysheva, *Computation of Atomic Processes* ("Adam Hilger" Institute of Physics Publishing, Bristol – Philadelphia, 1997).
27. A. S. Davydov, *Quantum Mechanics* (Pergamon, 1965).
28. C. Ramsauer and R. Kollath, Ann. Phys. (Leipzig) **12** 529 (1932).
29. D. F. Golden, H. W. Bandel, and J. A. Salerno, Phys. Rev. **146** 40 (1966).
30. H. S. W. Massey and R. O. Ridley, Proc. Phys. Soc. (London) **A69** 659 (1956).
31. J. C. Tully and R. S. Berry, J. Chem. Phys. **51** 2056 (1969)
32. Yu. N. Demkov and V. S. Rudakov, Soviet Phys. JETP, **32**, 1103 (1971).
33. S. Motoki *et al*, J. Phys. B **33**, 4193 (2000).
34. Yu. N. Demkov and V. N. Ostrovskii, *Zero-Range Potentials and Their Application in Atomic Physics* (New York: Plenum, 1988).
35. H. Nishimura, A. Danjo and H. Sugahara, J. Phys. Soc. of Japan, **54**, 1757 (1985).

Appendix A

Different from zero and unity, the coefficients $a_{ij}$ at unknown $D_{lm}^{\pm}$ are given in the Table 1, where the following designations are used: $C_0 = -2ikf_0$, $C_1 = -2ikf_1$, $Y_{lm} \equiv Y_{lm}(\mathbf{R}) \equiv Y(\vartheta_R, \varphi_R)$; $h_l = h_l(\rho)$; $h_l' = dh_l/d\rho$; $\rho = kR$.

Table 1. The coefficient $a_{ij}$

| | | | |
|---|---|---|---|
| $a_{12} = C_0\sqrt{\pi}h_0 Y_{00}$ | $a_{16} = -C_0 4\pi h_1 Y_{00} Y_{1-1}$ | $a_{17} = -C_0 4\pi h_1 Y_{00} Y_{10}$ | $a_{18} = -C_0 4\pi h_1 Y_{00} Y_{11}$ |
| $a_{21} = a_{12}$ | $a_{23} = -a_{16}$ | $a_{24} = -a_{17}$ | $a_{25} = -a_{18}$ |
| $a_{32} = -C_1\sqrt{\pi}h_1 Y_{11}$ | $a_{36}$, see below | $a_{37}$, see below | $a_{38}$, see below |
| $a_{42} = C_1\sqrt{\pi}h_1 Y_{10}$ | $a_{46}$, see below | $a_{47}$, see below | $a_{48} = -a_{37}$ |
| $a_{52} = -C_1\sqrt{\pi}h_1 Y_{1-1}$ | $a_{56}$, see below | $a_{57} = -a_{46}$ | $a_{58} = a_{36}$ |
| $a_{61} = -a_{32}$ | $a_{63} = a_{36}$ | $a_{64} = a_{37}$ | $a_{65} = a_{38}$ |
| $a_{71} = -a_{42}$ | $a_{73} = a_{46}$ | $a_{74} = a_{47}$ | $a_{75} = a_{48}$ |
| $a_{81} = -a_{52}$ | $a_{83} = a_{56}$ | $a_{84} = a_{57}$ | $a_{85} = a_{58}$ |

$a_{36} = C_1\{-4\pi Y_{11}Y_{1-1}h_1' + (h_1/\rho)[\sqrt{4\pi/5}Y_{20} + \sqrt{16\pi}Y_{00}]\}$;

$a_{37} = C_1\{-4\pi Y_{11}Y_{10}h_1' + (h_1/\rho)\sqrt{12\pi/5}Y_{21}\}$;

$a_{38} = C_1\{-4\pi Y_{11}Y_{11}h_1' + (h_1/\rho)\sqrt{24\pi/5}Y_{22}\}$;

$a_{46} = C_1\{4\pi Y_{10}Y_{1-1}h_1' - (h_1/\rho)\sqrt{12\pi/5}Y_{2,-1}\}$;

$a_{47} = C_1\{4\pi Y_{10}Y_{10}h_1' - (h_1/\rho)[\sqrt{16\pi/5}Y_{20} - \sqrt{16\pi}Y_{00}]\}$;

$a_{56} = C_1\{-4\pi Y_{1-1}Y_{1-1}h_1' + (h_1/\rho)\sqrt{24\pi/5}Y_{2,-2}\}$. \hfill (A1)

For the calculation of the functions $a_{\lambda\mu}$ resulting from the action of operators $\hat{B}_{lm}^{\pm}$ on a product of the functions in the square brackets in Eqs. (4) and (5) the following formulas are used:

$$\hat{B}_{lm}^{\pm}G_k(\mathbf{r}, \mp\mathbf{R}/2) = \hat{B}_{lm}^{\pm}\frac{e^{ik|\mathbf{r}\pm\mathbf{R}/2|}}{2\pi|\mathbf{r}\pm\mathbf{R}/2|} = P_{kl}(|\mathbf{r}\pm\mathbf{R}/2|)Y_{lm}(\mathbf{r}\pm\mathbf{R}/2). \hfill (A2)$$

$\hat{B}_{00}^{\pm} = \sqrt{\pi}$;

$\hat{B}_{1-1}^{\pm} = \mp\sqrt{6\pi}\frac{1}{k}\left(\frac{\partial}{\partial R_x} - i\frac{\partial}{\partial R_y}\right) = \mp\sqrt{12\pi}\frac{1}{k}\nabla_{-1}$;

$\hat{B}_{10}^{\pm} = \mp\sqrt{12\pi}\frac{1}{k}\frac{\partial}{\partial R_z} = \mp\sqrt{12\pi}\frac{1}{k}\nabla_0$;

$\hat{B}_{11}^{\pm} = \pm\sqrt{6\pi}\frac{1}{k}\left(\frac{\partial}{\partial R_x} + i\frac{\partial}{\partial R_y}\right) = \mp\sqrt{12\pi}\frac{1}{k}\nabla_{+1}$. \hfill (A3)

The results of operating by the cyclical components of the operator $\nabla_{0,\pm1}$ on the spherical functions $Y_{lm} \equiv Y_{lm}(\mathbf{R})$ are described by the following formulas [24]

$$\nabla_{-1}Y_{lm} = -\frac{l}{R}\sqrt{\frac{(l-m+1)(l-m+2)}{2(2l+1)(2l+3)}}Y_{l+1m-1} - \frac{l+1}{R}\sqrt{\frac{(l+m-1)(l+m)}{2(2l-1)(2l+1)}}Y_{l-1m-1};$$

$$\nabla_{+1}Y_{lm} = -\frac{l}{R}\sqrt{\frac{(l+m+1)(l+m+2)}{2(2l+1)(2l+3)}}Y_{l+1m+1} - \frac{l+1}{R}\sqrt{\frac{(l-m-1)(l-m)}{2(2l-1)(2l+1)}}Y_{l-1m+1};$$

$$\nabla_{0}Y_{lm} = -\frac{l}{R}\sqrt{\frac{(l+1)^2-m^2}{(2l+1)(2l+3)}}Y_{l+1m} + \frac{l+1}{R}\sqrt{\frac{l^2-m^2}{(2l-1)(2l+1)}}Y_{l-1m}. \tag{A4}$$

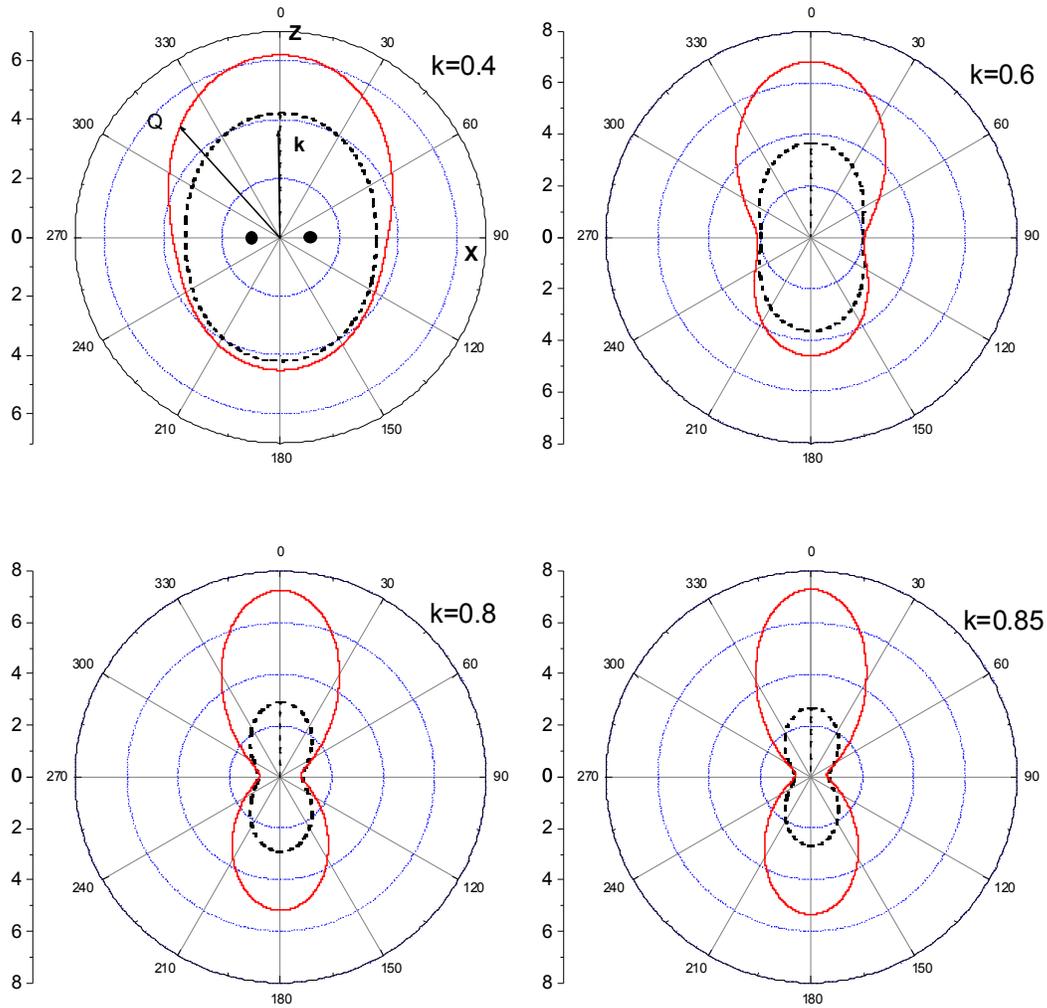

Fig. 1: Differential cross sections for electron elastic scattering by $C_2$ molecule (in au) for different electron energies $\varepsilon = k^2/2$. Solid circles in the upper left panel are the schematic picture of carbon atoms. Similarly, the atoms are also on the other panels, i.e. the vector **R** connecting the molecular atoms is perpendicular to the vector **k**∥**Z**. Dashed black curves are the cross sections calculated with *s*-phase shift only; solid red curves are the results calculated with *s+p*-phase shifts.

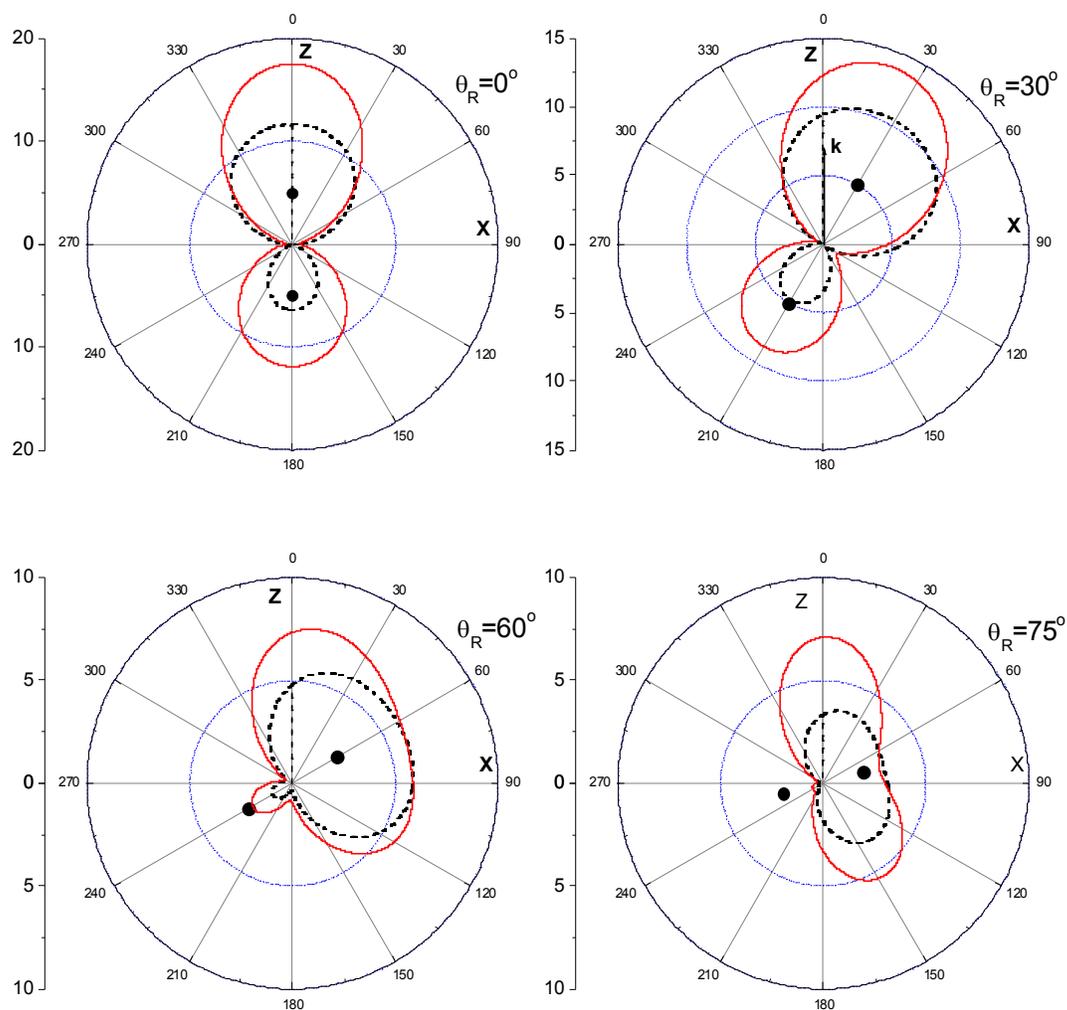

Fig. 2: Differential cross sections (in au) for electron elastic scattering ($k=0.8$ au) by $C_2$ molecule for different polar angles $\vartheta_R$ between the vectors **R** and **k**. Vectors **k**, **k'** and **R**, as in Fig. 1, are in the plane of the page. Solid circles in the panels are the schematic picture of carbon atoms. Dashed black curves are the cross sections calculated with *s*-phase shift only; solid red curves are results calculated with *s+p*-phase shifts.

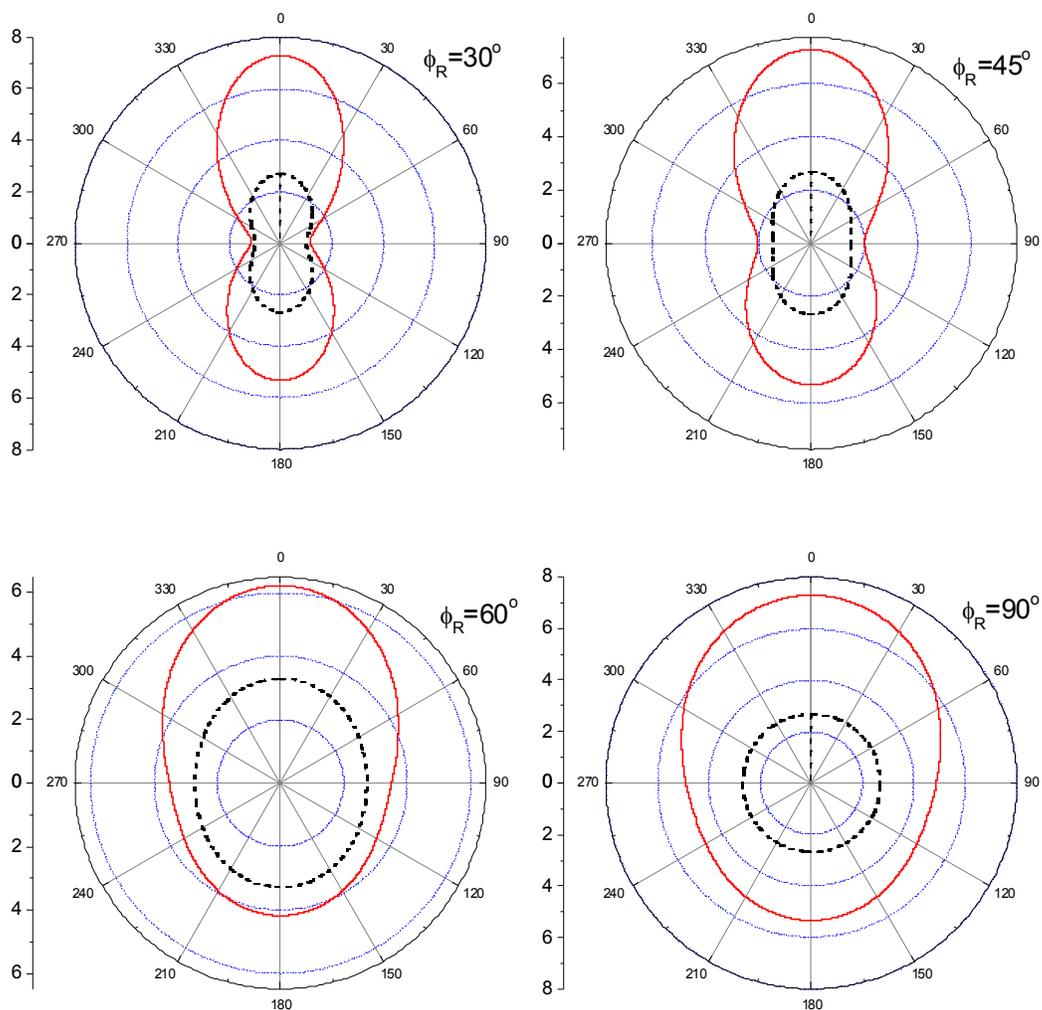

Fig. 3: Differential cross sections (in au) for electron elastic scattering ($k$=0.85 au) by $C_2$ molecule for different azimuthal angles $\varphi_R$ and fixed polar angle $\vartheta_R = 90^o$ between the vectors **R** and **k**. Dashed black curves are the cross sections calculated with *s*-phase shift only; solid red curves are results calculated with *s*+*p*-phase shifts.

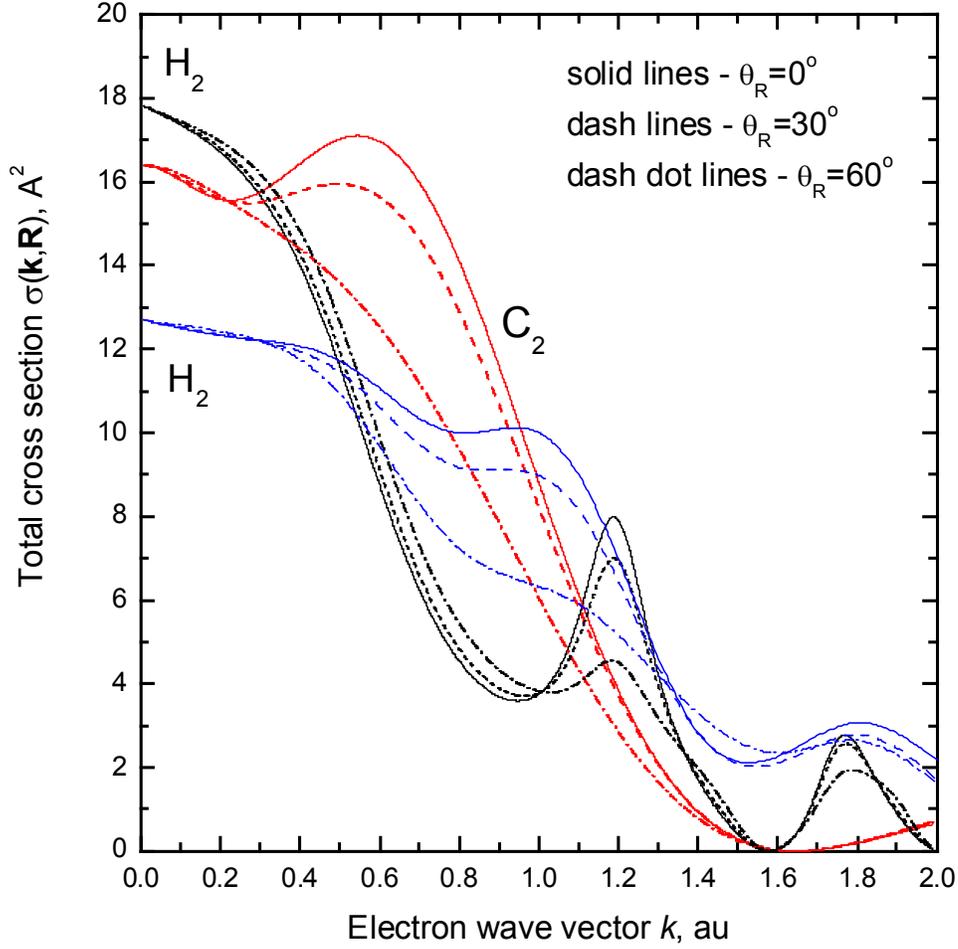

Fig. 4: Total cross sections for electron elastic scattering by fixed-in-space $H_2$ and $C_2$ molecules (in $Å^2$). The set of curves for $H_2$ molecule (with the beginning at 17.80$Å^2$) is calculated with Eq. (16) using $\delta_0^t(k)$ phase shift for single H atom. The set of curves for $H_2$ molecule (with the beginning at 12.70$Å^2$) is calculated with Eq. (17) using $\delta_0^t(k)$ and $\delta_1(k)$ for single H atom. The set of curves for $C_2$ molecule (with the beginning at 16.40$Å^2$) is calculated with Eq. (17) using $\delta_0(k)$ and $\delta_1(k)$ for single C atom. Solid, dash and dash dot lines (black and blue for $H_2$ and red for $C_2$ molecules) are the cross sections for different angles between the vectors $\mathbf{k}$ and $\mathbf{R}$; these angles are $\vartheta_R = 0^o$, $30^o$ and $60^o$, respectively.

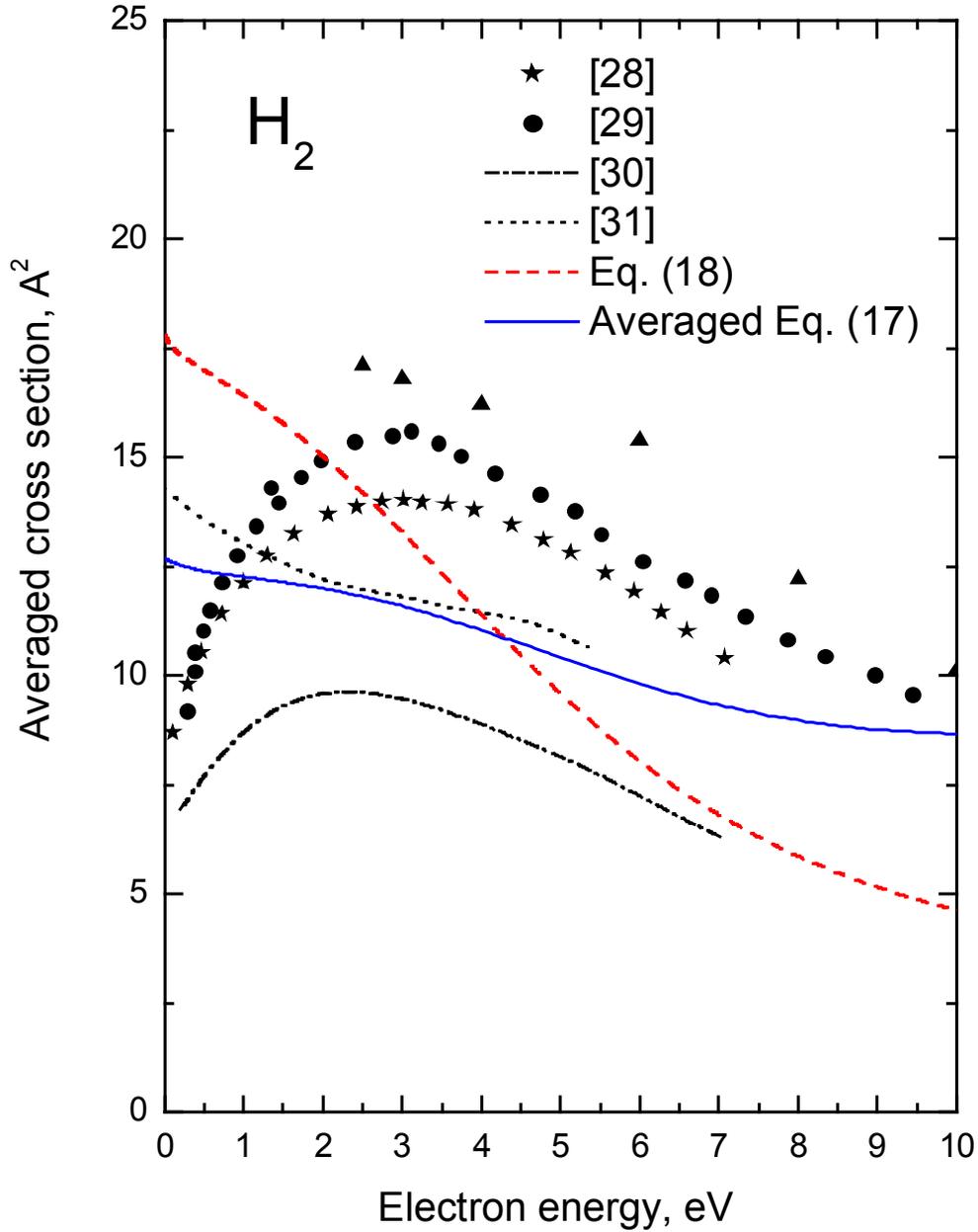

Fig. 5: Total cross sections $\bar{\sigma}(\varepsilon)$ for electron elastic scattering by $H_2$ molecules (in Å$^2$) as a function of electron energy $\varepsilon$. Black stars and circles are experiments; black triangles are experimental data from Table 2; dash dot line and dotted line are results of calculations; dash red line is our present calculations of $\bar{\sigma}(\varepsilon)$ with Eq. (18); solid blue line is our result of averaging (over the solid angles between the vectors **k** and **R**) the cross sections $\sigma_{sp}(\mathbf{k},\mathbf{R})$ that was preliminarily calculated with Eq. (17).